\documentclass[12pt]{article}
\usepackage[utf8]{inputenc}
\usepackage[russian, english]{babel}
\usepackage{amsmath}
\usepackage{amsfonts}
\usepackage{amssymb}

\usepackage{bm}

\renewcommand{\theequation}{\thesection.\arabic{equation}}

\def\lb{\label}
\def\be{\begin{equation}}
\def\ee{\end{equation}}

\begin{document}

\begin{titlepage}

\vspace*{0.7cm}

\begin{center}

{\LARGE\bf Generalized Wigner operators}

\vspace{0.3cm}



{\LARGE\bf and relativistic gauge fields}

\vspace{1.7cm}

{\large\bf I.L.\,Buchbinder$^{1,2,3}$\!\!\!, \, A.P.\,Isaev$^{2,4}$\!\!\!, \, M.A.\,Podoinitsyn$^{2}$\!\!, \,  S.A.\,Fedoruk$^2$}

\vspace{1.7cm}

\ $^1${\it Center for Theoretical Physics, Tomsk State Pedagogical University, Tomsk, 634041 Russia}, \\
{\tt joseph@tspu.edu.ru}

\vskip 0.5cm

\ $^2${\it Bogoliubov Laboratory of Theoretical Physics, JINR, 141980 Dubna, Moscow region, Russia}, \\
{\tt isaevap@theor.jinr.ru, mpod@theor.jinr.ru, fedoruk@theor.jinr.ru}

\vskip 0.5cm

\ $^3${\it Tomsk State University of Control Systems and Radioelectronics, 634034, Tomsk, Russia}

\vskip 0.5cm

\ $^4${\it Lomonosov Moscow State University, Physics Faculty, Russia} \\

\end{center}

\vspace{0.5cm}

\selectlanguage{english}
\begin{abstract}
We introduce and study the generalized Wigner operator. By
definition, such an operator transforms the Wigner wave function
into a local relativistic field corresponding to an irreducible
representation of the Poincar\'e group by extended discrete transformations, with integer
helicities $\lambda$ and $-\lambda$. It is shown that the
relativistic fields constructed in this way are gauge potentials and
satisfy the relations that determine free massless higher spin
fields.
\end{abstract}

\selectlanguage{russian}

\vspace{5cm}



\vspace{1cm}

\end{titlepage}

\setcounter{footnote}{0}
\setcounter{equation}{0}

\newpage

\setcounter{equation}0
\section{Introduction}

Space-time symmetries play a fundamental role in modern theoretical
high-energy physics. The most important of these symmetries is
formulated in terms of representations of the Poincar\'e group. The
description of unitary irreducible representations of the
four-dimensional Poincar\'e group was first given in the pioneering
works of Wigner \cite{Wig39}, \cite{Wig48} and Bargmann-Wigner
\cite{BarWig48}.

As well known that the physically interesting representations of the
Poincar\'e group are divided into massive and massless ones
depending on the value of the second-order Casimir operator
calculated on these representations. Massless representations in
turn are subdivided into representations with definite finite
helicities and representations with infinite (continuous) spin.
There is an extensive literature devoted to massive and massless
representations with finite helicity (see, for example, 
\cite{FG}-\cite{TUNG} and references therein). The
representations of infinite spin, which are formally physically
acceptable, did not attract mach attention in field theory for a
long time. However, in the last two decades there has been a surge
of interest in the construction of the Lagrangian theory of infinite spin fields
(see for e.g. the review \cite{BekSk} and
references therein).

Starting with the Wigner's paper \cite{Wig39} and the subsequent
papers \cite{Wig48}, \cite{BarWig48}, it is known that unitary
irreducible representations of the Poincar\'e group are induced by
unitary irreducible representations of a small group that preserves
the fixed momentum of a relativistic particle. The functions, on
which these representations are realized, are called Bargmann-Wigner (or Wigner) wave functions. 
The transitions of the Wigner wave functions to relativistic local fields are
given by the Wigner operators. In the case of a massless particle in
four dimensions, the small group is $ISO(2)$, which is the group of
motions of two-dimensional Euclidean space (a description of its
representations is given, for example, in \cite{Vilen}-\cite{ZhSh}).
It is known that unitary $ISO(2)$ representations are characterized by one
dimensionless positive real parameter, which we will denote as
$\bm{\rho}$. Such representations are realized on functions\footnote{In
an alternative formulation, which we will also use here, the
representation is given on the Fourier components of such functions
(see details in \cite{BIPF}).} defined on a circle of radius
$\bm{\rho}$. The representation space of the small group is
finite-dimensional for $\bm{\rho}=0$ and infinite-dimensional for
$\bm{\rho} \neq 0$.

In our previous paper \cite{BIPF}, two types of local relativistic
fields were constructed on which massless unitary irreducible
representations of the $4D$ Poincar\'e group with infinite spin are
realized. These fields are defined on a space parameterized by a
4-momentum and an auxiliary commuting variable: vector or spinor. In
constructing such field representations, in \cite{BIPF}, a
generalization of the Wigner operator was given, the kernel of which
defines an integral transform of the Wigner wave functions into
local relativistic fields.

As mentioned above, another example of irreducible representations
of the Poincar\'e group is massless representations with definite
helicity. In this paper, we will show that the generalized Wigner
operator, introduced in \cite{BIPF} for infinite spin
representations, can also be defined for constructing the
conventional massless local relativistic fields corresponding to
helicity representations. Here, for simplicity, we will consider
only the case of integer helicities.\footnote{Some similar
constructions for fields of half-integer helicities were considered
in \cite{ZFed}.}

It is well known that the irreducible massless field representations
of the Poincar\'e group with definite helicities described in the
literature are given in terms of spin-tensor fields, which, in the
terminology of field theory, are called field strengths (see, for example,
\cite{Weinb1},\,\cite{Weinb2}, \,\cite{Nov},\,\cite{BK} and
references therein). At the same time, in Lagrangian field models, the
fundamental objects are potentials, not the strengths. Therefore, it
seems interesting to describe the representations of the Poincar\'e
group directly in terms of potentials. However, the following
circumstance must be borne in mind here. In $P$-parity invariant
Lagrangian massless theories, the fields (potentials) always contain
two helicities $\lambda$ and $-\lambda$. From the proper
Poincar\'e group point of view, this means that the corresponding fields
(potentials) refer to the reducible representation. Nevertheless,
the two helicities $\lambda$ and $-\lambda$ can be included into one
irreducible representation of the extended Poincar\'e group
including in general case both transformations continuously related
to identity and discrete $C$, $P$, $T$ transformations (see, for
example, \cite{FG},\,\cite{TUNG} and especially the papers
\cite{GS},\,\cite{BGS}, which are specially devoted to the
description of discrete transformations in the Poincar\'e group).
Therefore, in what follows, by an irreducible representation of a
Poincar\'e group we mean representations of the extended Poincar\'e
group.

In this paper, following \cite{BIPF}, we introduce the generalized
Wigner operator corresponding to the case $\bm{\rho} = 0$, derive
equations for the kernel of this operator, and find their solution. This solution determines the Wigner operator that transforms the Wigner wave
functions into Lorentz-covariant local fields corresponding to massless
particles with certain helicities. In the resulting scheme, if one fixes $\rho = 0$, the group $ISO(2)$ is reduced
to the group $SO(2)$ and we obtain that unitary irreducible representations
of the group $ISO(2)$ decompose into an infinite direct sum of one-dimensional
representations of the group $SO(2)$ that
induces massless helical representations of the Poincar\'e group. To
construct irreducible representations of the extended Poincar\'e
group that are of interest to us, we combine the representations
with $\lambda$ and $-\lambda$ helicities into one irreducible
representation of the extended group. It turns out that the
constructed local relativistic fields are defined up to a gauge
transformation and correspond to free massless higher spin fields.

The work is organized as follows. Section\,\ref{R1} briefly
discusses the basic notions related to the description of unitary
irreducible massless helical representations of the Poincar\'e group
in terms of the Wigner wave function. Section\,\ref{R2} is devoted
to the construction of a generalized Wigner operator for
representations of the type under consideration. In
section\,\ref{R3}, equations are constructed for the integral kernel
of the generalized Wigner operator, which makes it possible to
construct relativistic fields including two helicities $\lambda$ and
$-\lambda$. In section\,\ref{R4}, it is shown that the relativistic
field generated by the generalized Wigner operator is defined up to
a gauge transformation and satisfies the conditions that determine
free massless higher spin fields. In the Conclusion, the main
results of the work are briefly formulated and some directions for
further study are noted. The details of the calculations used in the
the paper are referred to the Appendix.

\setcounter{equation}{0}
\section{Wigner wave function} \lb{R1}

The construction of unitary irreducible massless $4D$
representations of the Poincar\'e covering group $ISL(2,\mathbb{C})$
was first carried out by Wigner \cite{Wig39}-\cite{Wig48} and
Bargmann-Wigner \cite{BarWig48} in terms of the functions
$\Phi(p,\varphi)$, later called Wigner wave functions. By
construction, these functions depend on the isotropic 4-momentum
$p_\mu$ satisfying the condition $p^\mu p_\mu=0$ and on the
additional coordinate $\varphi\in[0,2\pi)$. The action of an element $A\in
SL(2,\mathbb{C})$ on Wigner function $\Phi(p,\varphi)$ is defined 
by the following relation\footnote{The connection between the Poincar\'e group and its
covering, as well as the choice of the test momentum for the
stability subgroup and the definition of the Wigner operator in
2-dimensional spinor space, are presented in Appendix\,1.}:
\begin{eqnarray}
\nonumber
[U(A) \Phi](p,\varphi)  &=&  \left. e^{- i \vec{b} \cdot \vec{t}_{\varphi}} \Phi \bigl(\Lambda^{-1} p , \; \varphi - \theta \bigr)\right|_{\vec{b} = \vec{b}_{A,\Lambda^{-1} p}, \;
\theta = \theta_{A,\Lambda^{-1} p}}  \\ [7pt]
\lb{trans-Wvf-p}
&=& \int\limits_{0}^{2\pi} d \varphi^\prime \  \mathcal{D}_{\varphi\varphi^{\prime}}(\theta_{A,\Lambda^{-1} p}, \vec{b}_{A,\Lambda^{-1} p}) \,
\Phi (\Lambda^{-1} p , \varphi^{\prime})\,,
\end{eqnarray}
where $\Lambda\in SO(1,3)$ is associated with $A$ by the transformation
\be
\lb{Alambd-p}
A\, \sigma^\mu A^{\dagger} = \sigma^\nu \; \Lambda_\nu^{\;\; \mu}(A)\,, \;\;\; A
\in SL(2,\mathbb{C})\, , \;\;\; \Lambda_\nu^{\;\; \mu}(A) \in O(1,3)
\,.
\ee
Here $\mathcal{D}_{\varphi\varphi^{\prime}}(\theta_{A \Lambda^{-1} p}, \vec{b}_{A,\Lambda^{-1} p})$ is the matrix element of 
the unitary irreducible representation of the group $ISO(2)$, whose parameters $\theta$ and $\vec{b}$ depend on the Lorentz transformations and 4-momentum (see the relations (\ref{acgf1-p}) and (\ref{acgf1-p1}) in Appendix\,1). This matrix element is written as
\be \lb{D-a-p}
\mathcal{D}_{\varphi \varphi^{\prime}}(\theta, \vec{b}) =
e^{  - i \, \vec{b}_{\mathstrut} \cdot  \vec{t}_{\varphi}}\delta(\varphi-\varphi^{\,\prime}-\theta) \, ,
\ee
where $\delta(\varphi)$ is the periodic $\delta$-function, the 2-vector $\vec{t}_{\varphi}$ has components $||\vec{t}_{\varphi}| | = \bm{\rho} (\cos \varphi, \sin \varphi)$, and the positive real parameter $\bm{\rho}$ characterizes the unitary irreducible representation of the $ISO(2)$ group.

In the case of $\bm{\rho}\neq0$, the representations of $ISO(2)$ are
infinite-dimensional and irreducible, while for $\bm{\rho}=0$ we have $\vec{t}_\varphi=0$ and in this representation the group $ISO(2)$ is reduced to its subgroup $SO(2 )$. In such a case, the representation of
the small group entering into relation (\ref{trans-Wvf-p}) is decomposed into a direct sum of one-dimensional representations. The easiest way to see this is to write massless representations
(\ref{trans-Wvf-p}) in terms of the Fourier components $\Phi_{n}(p)$, $n\in \mathbb{Z}$ 
in the Fourier expansion of $\Phi(p,\varphi)$:
\be \lb{Phi-exp-p}
\Phi(p,\varphi)
= \sum_{n=-\infty}^{\infty} \Phi_{n}(p)
 e^{i n \varphi} \,.
\ee
In such a discrete representation, the transformation (\ref{trans-Wvf-p}) becomes
\be \lb{rwd2-p}
[U(A) \Phi]_{n}(p) =  \sum_{m=-\infty}^{\infty}  \mathcal{D}_{nm} (\theta_{A,\Lambda^{-1} p},
\vec{b}_{A,\Lambda^{-1} p}) \, \Phi_m (\Lambda^{-1} p) \, ,
\ee
where $\mathcal{D}_{nm}$ is the matrix of the element $h$ of the little group in the discrete basis:
\be \lb{mn-b-p}
 \mathcal{D}_{nm} ( \theta, \vec{b}) =
 (- i e^{i \beta})^{m-n} \, e^{- i m \theta} J_{(m-n)}(b{\bm{\rho}}) \; ,
\ee
the real numbers $\beta$ and $b$ are the polar coordinates of the 2-vector
$\vec{b} = b \, (\cos \beta, \sin \beta)$ and $J_{(n)}(x)$ are the Bessel functions of integer order.
In the case of ${\bm{\rho}} \to 0$ taking into account the property $J_{n-m} (0) = \delta_{nm}$ for
the Bessel functions, the matrix element (\ref{mn-b-p}) is written as
\be \lb{rwd1-b}
\mathcal{D}_{nm}(\theta,\vec{b})=\delta_{nm}e^{-in\theta} \; ,
\ee
i.e. the matrix $\mathcal{D}(\theta,\vec{b})$ becomes diagonal and the transformation (\ref{rwd2-p}) 
is written as
\be \lb{rwd1-a}
[U(A) \Phi]_{n}(p) = e^{-  i n \theta} \, \Phi_n (\Lambda^{-1} p) \, ,
\ee
which also immediately follows from the first equality in (\ref{trans-Wvf-p}).
Thus, in the case ${\bm{\rho}} \to 0$, the little group $ISO(2)$ on the representation (\ref{trans-Wvf-p}) is
reduced to its subgroup $SO(2)$ and the representation of the little group is decomposed into the sum of an
infinite number of one-dimensional irreducible representations. For a fixed $n$, a chain of inducings takes
place: the one-dimensional irreducible representation $SO(2)$ induces an irreducible unitary representation $ISO(2)$
with a trivial realization of two-dimensional translations, which in turn induce the helical representations of
the Poincar\'e group $ISO(1,3)$.

\setcounter{equation}{0}
\section{Generalized Wigner operator} \lb{R2}

The main objects of the relativistic Lagrangian field theory are the
space-time local spin-tensor fields characterized by standard
transformations with respect to the covering of the Poincar\'e group
$ISL(2,\mathbb{C})$. At the same time, the Wigner wave functions
$\Phi(p,\varphi)$ are defined in the momentum space and
transformed under $ISL(2,\mathbb{C})$ action according to (\ref{trans-Wvf-p}) or (\ref{rwd2-p}).
After Fourier transformations to a coordinate representation, 
these Wigner wave functions do not coincide with any
space-time fields used in the Lagrangian field theories. The transition
from the description of irreducible representations of
$ISL(2,\mathbb{C})$ in terms of Wigner wave functions to the
description in terms of local fields with finite spin is carried out
by means of the Wigner operators. Such a transition for massive
and massless representations was proposed by Weinberg \cite{Weinb1},
\cite {Weinb2}\footnote{See also \cite{IsPod0} for analysis of
massive representations.}.

In our recent paper \cite{BIPF}, we applied the method of induced Wigner
representations to describe irreducible representations
of the Poincar\'e group of infinite spin and constructed the generalization of
the Wigner operators. Our approach is closed to the approach
developed earlier in \cite{ShTor1}, \cite{ShTor2}. Here we apply
the generalized Wigner operator to construct
local fields of integer helicities. It turns out that in this case such
an approach automatically leads to massless fields with gauge
symmetry.

Following \cite{BIPF}, we consider irreducible
representations realized in the space of fields $\Psi(p, \eta)$,  where
$p$ is the space-time 4-momentum $p^\mu$ and $\eta$ is an additional
commuting vector variable $\eta^\mu$. 
The Fourier transferring the fields
$\Psi(p, \eta)$ to the coordinate representation and expanding the
result in $\eta$ we should obtain the local spin-tensor fields
\footnote{In \cite{ZFed}, \cite{GS}, \cite {BGS} the irreducible
representations were constructed using the additional spinor
variables.}.
It is natural to assume \cite{Weinb1}, \cite{Weinb2} that the relativistic fields
$\Psi(p, \eta)$ are in one-to-one correspondence with the Wigner wave functions
$\Phi(p,\varphi)$ which are transformed under Poincar\'e group action according to
(\ref{trans-Wvf-p}). This one-to-one correspondence is given by the
integral transform
\be
 \Psi(p, \eta) \ = \ \int\limits_{0}^{2\pi} d \varphi \, \mathcal{A}(p,\eta,\varphi)\, \Phi (p, \varphi) \, .
\ee
The integral operator with the kernel
$\mathcal{A}(p,\eta,\varphi)$ is called the generalized Wigner
operator. In a discrete basis, the transition from the Wigner wave
functions $\Phi_n(p)$ to the local fields $\Psi(p, \eta)$ takes the
form
\be
\lb{rwd3} \Psi(p,\eta) = \sum_{n=-\infty}^{\infty}
\mathcal{A}(p,\eta,n) \Phi_n (p) \, ,
\ee
where the kernel $\mathcal{A}(p,\eta,n)$ of the
generalized Wigner operator in the discrete basis
 is the Fourier component of the kernel
$\mathcal{A}(p,\eta,\varphi)$: \be \lb{rwd4} \mathcal{A}(p,\eta,n) =
\int\limits_{0}^{2\pi} d \varphi  \, \mathcal{A}(p,\eta,\varphi) \,
e^{i n \varphi} \, . \ee

The requirement that the field $\Psi(p, \eta)$ be local is a very
strong condition. It assumes that the transformation of the field
$\Psi(p, \eta)$ (in momentum space) under the Lorentz
transformations $A\in SL(2, \mathbb{C})$ has the same form as the
corresponding transformation of any local relativistic field:
\be
\lb{unpgr41-p}
[U(A) \Psi](p, \eta) \ = \
 \Psi\bigl(\Lambda^{-1}(A)\, p , \; \Lambda^{-1}(A) \, \eta
 \bigr) \, ,
\ee where the matrices $\Lambda$ and $A$ are linked by
(\ref{Alambd-p}). Substituting (\ref{trans-Wvf-p}) and (\ref{rwd3})
into (\ref{unpgr41-p}) leads to a transformation rule for the kernel of the
generalized Wigner operator $\mathcal{A}(p,\eta, n)$ in a discrete
basis:
\be
\lb{rwd44}
\mathcal {A} (\Lambda^{-1} p, \; \Lambda^{-1}
\eta, n) = \sum_{m=-\infty}^{\infty} \mathcal{A} (p, \eta, m)\,
\mathcal{D}_{mn} (\theta_{A, \; \Lambda^{-1} p},
\vec{b}_{A,\Lambda^{-1} p}) \, .
\ee
The equation (\ref{rwd44}) relates the
transformation of the kernel $\mathcal {A}$ under the
4-dimensional Lorentz transformations (left-hand side) with the
transformations of $\mathcal {A}$ under the action of the little
group on it (right-hand side).

Relation (\ref{rwd44}) gives to two important equations. First,
take in (\ref{rwd44}) as $A = A_{(p)}$ and $\Lambda = \Lambda(A_{(p)})$,
where the matrix $A_ {(p)} \in SL(2,\mathbb{C})$ is defined in Appendix 1.
Then, taking into account (\ref{add-re-p}), for (\ref{rwd44}) we obtain
\be
\lb{unpgr8-p}
\mathcal{A}(\overset{_{\mathrm{\;o}}}{p}, \Lambda^{-1}(A_{(p)})
\eta, n)
 \ = \  \mathcal{A}(p,\eta,n) \, .
\ee
It means that to construct the generalized Wigner operator
$\mathcal{A}(p,\eta,n)$ for arbitrary 4-momentum $p^\mu$, it
suffices to know the form of this operator $\mathcal{A} (\overset{_{
\mathrm{\;o}}}{p}, \eta, n)$ for the test momentum
$\overset{_{\mathrm{\;o}}}{p}{}^\mu$. Second, consider the
expression (\ref{rwd44}) with $p = \overset{_{\mathrm{\;o}}}{p}$ and
take $\Lambda = \Lambda(h)$ , where $h$ is an arbitrary element
of the stability subgroup
$\overset{_{\mathrm{\;o}}}{p}{}^\mu$ of the test momentum ($h$ depends on the parameters
$\theta$ and $\vec{ b}$). As a result, relation (\ref{rwd44}) is converted to
\be
\lb{unpgr9-p}
\mathcal{A}(\overset{_{\mathrm{\;o}}}{p}, \Lambda^{-1}(h) \eta, n) =
\sum_{m=-\infty}^{\infty} \mathcal{A} (\overset{_{\mathrm{\;o}}}{p},
\eta, m)\,  \mathcal{D}_{mn} (\theta, \vec{b}) \, .
\ee
Here the
matrix $\Lambda(h)$ depends on the parameters $\theta$, $\vec{b}$
and is defined according to (\ref{Alambd-p}) by the relation $h \,
\sigma^\mu \, h^ \dagger = \sigma^\nu \, \Lambda_\nu^{\; \mu}(h)$,
where the matrix $h$ is given in (\ref{hhhh}).

Relation (\ref{unpgr9-p}) in the infinitesimal form with
small parameters $\theta$ and $\vec{b}$
leads to three differential equations on the function
$\mathcal{A}(\overset{_{\mathrm{\;o}}}{p},\eta,n)$. After solving
these equations and finding $ \mathcal{A}({p},\eta,n)$ using
(\ref{unpgr8-p}), we can construct, using relation (\ref{rwd3}), a
local relativistic field in momentum representation.

Considering the massless representations of the infinite integer spin
in \cite{BIPF}, we obtained two types of solutions for the
relativistic field, which were called singular and non-singular
solutions in \cite{ShTor1}, \cite{ShTor2}. In the following
sections, we will show how, within the framework of our approach,
one can construct the standard massless helical relativistic fields.

\setcounter{equation}{0}
\section{Finding the generalized Wigner operator} \lb{R3}

Let us move on  to the construction of massless relativistic fields of finite helicities.
First of all, we emphasize the important points
of our approach in the construction of
local fields in terms of irreducible representations of the
Poincar\'e group.

First, in order to relate our approach to the Lagrangian field
theory, it is natural to assume that the constructed relativistic
fields must be spin-tensor fields (purely tensor fields in the case
of integer helicities), which are gauge potentials for non-zero
helicities, i.e. defined up to transformations with local
parameters. Standard massless local fields which realize irreducible
massless representations of the Poincar\'e group with non-zero
integer helicities, are field strengths
(see e.g. \cite{Weinb1}, \cite{Weinb2}, \cite{Nov},
\cite{BK}), not potentials. However, the Lagrangian models of
relativistic field theory are usually formulated in terms of
potentials. Thus, the description of irreducible representations in
terms of potentials and their equations of motion for the purposes of field theory seem to be more
suitable or even more fundamental than the description in terms of
field strengths.

Second, in field theories with unbroken parity symmetry, the field
description of massless states with the help of potentials uses
states with both helicities $\lambda$ and $-\lambda$, which
corresponds to irreducible representations of the extended
Poincar\'e group including discrete $C$, $P$, $T$
transformations\footnote{From the point of view of the proper
Poincare group, such representations are reducible. In the extended
Poincare group, they become irreducible due to discrete
transformations relating fields with helicity $\lambda$ and
$-\lambda$.}. For example, the vector-potential $A_\mu$ of the the
electromagnetic field describes states with helicities $\pm1$, while
states with a fixed helicity are described by the components
$F_{\alpha\beta}\sim (\sigma^{\mu\nu})_{\alpha\beta}F_{\mu\nu}$ and
$\bar F_{\dot\alpha\dot\beta}\sim
(\tilde\sigma^{\mu\nu})_{\dot\alpha\dot\beta}F_{\mu\nu}$ of the one
field strengths $F_{\mu\nu}=\partial_\mu A_\nu-\partial_\nu A_\mu$.
We show below that the parameter $n$ in the Wigner wave function
(\ref{rwd1-a}) coincides with the helicity. Therefore, to describe
the relativistic fields in our approach, we will use two Wigner wave
functions $\Phi_{\pm n} (p)$ for a fixed $n$, except the scalar case
$n=0$. In other words, to construct a relativistic field with
helicities $\lambda=\pm n$ in expansion (\ref{rwd3}), it is
necessary to take into account only two terms with $(\pm n)$. Thus,
the relativistic field containing states with helicities
$\lambda=\pm{n}$ has the form:
\be
\lb{rfield}
\Psi_n(p,\eta) \ = \
\mathcal{A}(p,\eta,n) \Phi_n (p)  \ +  \ \mathcal{A}(p,\eta,-n)
\Phi_{-n} (p) \, .
\ee

In the case under consideration, which corresponds to
finite-dimensional representations of the little group $ISO(2)$,
i.e. for ${\bm{\rho}} \,{=}\, 0$, the transformation law
(\ref{unpgr9-p}) in infinitesimal form leads to three equations:
\be
\lb{th-0}
 \left( \zeta \frac{\partial}{\partial \zeta} - \bar{\zeta} \frac{\partial}{\partial \bar{\zeta}} \right) \mathcal{A}(\overset{_{\mathrm{\;o}}}{p},\eta,n) \ = \ n \, \mathcal{A}(\overset{_{\mathrm{\;o}}}{p},\eta,n)\,.
\ee
\be \lb{b-0}
\left\{
\begin{array}{rcl}
\displaystyle
 \left(\zeta \frac{\partial}{\partial \eta^{-}} + \eta^{+} \frac{\partial}{\partial \bar{\zeta} } \right) \mathcal{A}(\overset{_{\mathrm{\;o}}}{p},\eta,n) = 0 \, ,
\\ [10pt]
\displaystyle
 \left(\bar{\zeta} \frac{\partial}{\partial \eta^{-}} + \eta^{+} \frac{\partial}{\partial \zeta } \right) \mathcal{A}(\overset{_{\mathrm{\;o}}}{p},\eta,n) = 0 \, ,
\end{array}
\right.
\ee
where we introduce the following variables: 
\be
\lb{nev1-p}
\zeta := \eta^2 + i \eta^1 \, , \quad \bar{\zeta} :=
\eta^2 - i \eta^1 \, ; \qquad \eta^{\pm} := \eta^0 \pm \eta^3 \,.
\ee
The left-hand sides of equations (\ref{th-0}) and (\ref{b-0})
are obtained by expanding $\Lambda(h)$ along the directions of $4D$ generators of the
massless test momentum stability subgroup (see \cite{BIPF}), and the
right-hand sides are obtained from the explicit form (\ref{rwd1-b})
of the matrix element $\mathcal{D}_{mn} (\theta, \vec{b})$.

Equation (\ref{th-0}) determines the degree of homogeneity of the
function $\mathcal{A}(\overset{_{\mathrm{\;o}}}{p},\eta,n)$ as a
function of the complex variable $ \zeta$. Since $ ( \zeta\,
\partial/ \partial \zeta - \bar{\zeta}\, \partial / \partial
\bar{\zeta} ) (\zeta\bar\zeta)=0$ , the general solution of
the equations (\ref{th-0}) can be written as:
\be
\lb{s-th0}
\mathcal{A}(\overset{_{\mathrm{\;o}}}{p},\eta,n) = \left\{
\begin{array}{lll}
\displaystyle
\zeta^{n} \,g^{_{(+)}}_n(\eta^{+},\eta^{-}, \zeta \bar{\zeta})  \;\;\;\;\; {\mbox{with}}\;\; n\geq0\, ,
\\ [10pt]
\displaystyle
\bar{\zeta}^{-n} \, g^{_{(-)}}_n(\eta^{+},\eta^{-}, \zeta \bar{\zeta}) \;\;\; {\mbox{with}}\;\; n<0\, ,
\end{array}
\right.
\ee
where the main polynomial part in $\zeta$ and
$\bar\zeta$ is explicitly distinguished (this is convenient for what
follows), while $g^{_{(\pm)}}_{n}(\eta^{+} ,\eta^{-}, \zeta
\bar{\zeta})$ are the arbitrary functions of the arguments
$\eta^{\pm}$ and $\zeta \bar{\zeta}$.

Let us find now the solution of equations (\ref{b-0}). Multiplying
the first equation of the system (\ref{b-0}) by $\bar{\zeta}$, the
second one by $\zeta$ and considering their difference, we get,
taking into account (\ref{th-0}),
\be
\lb{gf-a}
\eta^+\mathcal{A}(\overset{_{\mathrm{\;o}}}{p},\eta,n)  = 0 \,
\ee
provided $n \neq 0$. In the class of distributions, equation
(\ref{gf-a}) is solved in the form
$\mathcal{A}(\overset{_{\mathrm{\;o}}}{p},\eta,n)\sim
\delta(\eta^+)...$ Substituting such a function
 $\mathcal{A}(\overset{_{\mathrm{\;o}}}{p},\eta,n)$ into the system of equations (\ref{b-0}) leads
to the conclusion that the dependence of this function on the
variable $\eta^-$ is realized in a form of  combination of $(\eta^{+}\eta^{-}
- \zeta \bar{ \zeta})=\eta^\mu\eta_\mu =: \eta \cdot \eta$. Thus, the general
solution of equations (\ref{th-0}) and (\ref{b-0}) is:
\be
\lb{s-th0a}
\mathcal{A}(\overset{_{\mathrm{\;o}}}{p},\eta,n) =
\left\{
\begin{array}{lll}
\displaystyle
\zeta^{n} \, \delta(\eta^+) \, f^{_{(+)}}_n(\eta \cdot \eta)  \;\;\;\;\; {\mbox{with}}\;\; n>0\, ,
\\ [10pt]
\displaystyle
{\bar{\zeta}}^{-n} \, \delta(\eta^+) \,  f^{_{(-)}}_n(\eta \cdot \eta) \;\;\; {\mbox{with}}\;\; n<0\, ,
\end{array}
\right.
\ee
where $f^{_{(\pm)}}_{n}(\eta \cdot \eta)$ are the arbitrary
functions of one real variable $\eta \cdot \eta$. The expansion of these
functions in the argument $\eta \cdot \eta$ leads to fields with infinite
degeneracy of the helicity spectrum. To eliminate the resulting
degeneracy, it is necessary to fix the functions
$f^{_{(\pm)}}_{n}(\eta \cdot \eta)$. We choose the simplest condition
$f^{_{(\pm)}}_{n}(\eta \cdot \eta)=1$. In this case, the generalized Wigner
operator of the helicity state is
\be
\lb{s-th0b}
\mathcal{A}(\overset{_{\mathrm{\;o}}}{p},\eta,n) = \left\{
\begin{array}{lll}
\displaystyle
\delta(\eta^+) \, {\zeta}^{n} \,   \;\;\;\;\; {\mbox{with}}\;\; n>0\, ,
\\ [10pt]
\displaystyle
\delta(\eta^+) \, {\bar{\zeta}}^{-n} \,   \;\;\; {\mbox{with}}\;\; n<0\, .
\end{array}
\right. \ee Then, in the test momentum frame
$\overset{_{\mathrm{\;o}}}{p}_\mu$, the relativistic gauge field
(\ref{rfield}) describing the helicity states has the form:
\be
\lb{rfields} \Psi_n(\overset{_{\mathrm{\;o}}}{p},\eta) \ = \
\delta(\eta^+)\Big[
F^{_{(+)}}_{n}(\overset{_{\mathrm{\;o}}}{p},\eta)  \ +  \
F^{_{(-)}}_{n}(\overset{_{\mathrm{\;o}}}{p},\eta) \Big] , \ee where
\be \lb{rfield-a} F^{_{(+)}}_{n}(\overset{_{\mathrm{\;o}}}{p},\eta)
\ = \  \zeta^{n}\, \Phi_n (\overset{_{\mathrm{\;o}}}{p}) \,, \qquad
F^{_{(-)}}_{n}(\overset{_{\mathrm{\;o}}}{p},\eta) \ = \
\bar\zeta{}^{n}\, \Phi_{-n} (\overset{_{\mathrm{\;o}}}{p}) \,
\ee
contain arbitrary functions $\Phi_n (\overset{_{\mathrm{\;o}}}{p})$
and $\Phi_{-n} (\overset{_{\mathrm{\;o}}}{ p})$.

The fields (\ref{rfields}) are realized in the space of functions
depending on an additional 4-vector $\eta^\mu$ in addition to the
4-momentum. Appendix 2 presents the realization of the generators of
the Poincar\'e group and the Pauli-Lubanski vector on such a space.
In particular, the expressions (\ref{pvPL2}) show that the Pauli-Lubanski 4-vector is
proportional on the
functions $\delta(\eta^+) F^{_{(+)}}_{n} (\overset{_{\mathrm{\;
o}}}{p},\eta)$ and $\delta(\eta^+) F^{_{(-)}}_{n}
(\overset{_{\mathrm{\;o}}}{p},\eta)$ to the 4-momentum: \be \lb{rwd14}
\overset{_{\mathrm{\;o}}}{W}_\mu \left [ \delta(\eta^+)
F^{_{(\pm)}}_{n} (\overset{_{\mathrm{\;o}}}{p},\eta) \right ] = \pm
n \, \overset{_{\mathrm{\;o}}}{p}_\mu \left [ \delta(\eta^+)
F^{_{(\pm)}}_{n} (\overset{_{\mathrm{\;o}}}{p},\eta) \right ] \,.
\ee
Thus, the distributions $\delta(\eta^+) F^{_{(+)}}_{n}
(\overset{_{\mathrm{\;o}}}{p},\eta) $ and $\delta(\eta^+)
F^{_{(-)}}_{n} (\overset{_{\mathrm{\;o}}}{p},\eta)$ describe the
massless states with $n$ and $-n$ helicities, respectively.

Fields and generalized Wigner operator for arbitrary momentum $p^{\mu}$ can
be found (using the relation (\ref{unpgr8-p})) from
generalized Wigner operator (\ref{s-th0b}) defined on the test momentum. To do
this, we use the following relations:
$\zeta=\overset{_{\mathrm{\,o}}}{\varepsilon}_{(+)}\cdot\eta$,
$\bar\zeta=\overset{_{ \mathrm{\,o}}}{\varepsilon}_{(-)}\cdot\eta$,
where $\overset{_{\mathrm{\,o}}}{\varepsilon}_{(\pm
)}:=\overset{_{\mathrm{\,o}}}{\varepsilon}_{(2)}\pm
i\overset{_{\mathrm{\,o}}}{\varepsilon}_ {(1)}$ and the 4-vectors
$\overset{_{\mathrm{\,o}}}{\varepsilon}_{(1)}$,
$\overset{_{\mathrm{\,o}}}{\varepsilon}_{(2)}$ are introduced in
Appendix\,3. Substituting (\ref{s-th0b}) into (\ref{unpgr8-p}), we
find the explicit form of the generalized Wigner operator of helicity
states for an arbitrary 4-momentum:
\be
\lb{wo.fs}
\mathcal{A}(p,\eta,n) = \left\{
\begin{array}{lll}
\displaystyle
\delta(\eta \cdot p) \, (\varepsilon_{(+)} \cdot \eta)^{n} \,   \;\;\;\;\; {\mbox{with}}\;\; n>0\, ,
\\ [10pt]
\displaystyle
\delta(\eta \cdot p) \, (\varepsilon_{(-)} \cdot \eta)^{-n} \,   \;\;\; {\mbox{with}}\;\; n<0\, ,
\end{array}
\right.
\ee
where the 4-polarization vectors $\varepsilon_{(\pm)}$ defined in (\ref{pv-p}) are used.

\setcounter{equation}{0}
\section{Local gauge fields} \lb{R4}

Let us move on to the construction of the irreducible representations in
terms of the local relativistic fields. We substitute the kernel of the
generalized Wigner operator (\ref{wo.fs}) into the
expression (\ref{rfield}) and define a relativistic field in the form
\be
\lb{rwd7}
\Psi_n(p,\eta) \ = \ \delta(\eta \cdot p) \,F_n(p,\eta) \,
, \ee where \be \lb{rwd8a} F_n(p,\eta) \ = \ F^{_{(+)}}_{n} (p,
\eta) \ + \ F^{_{(-)}}_{n} (p, \eta) \, , \qquad F^{_{(\pm)}}_{n}
(p, \eta) = (\varepsilon_{(\pm)} \cdot \eta)^{n} \, \Phi_{\pm
n}(p)\, .
\ee
Standard fields of definite helicity are obtained by
expanding $F^{_{(\pm)}}_{n} (p, \eta)$ into a series in the vector
variable $\eta^\mu$ (see below). Note that the field
$\Psi_n(p,\eta)$ defined in this way is the distribution due
to the presence of the $\delta$-function, while the fields
$F^{_{(\pm)}}_ {n} (p, \eta)$ are the conventional functions of the variable
$\eta$, which have the same degree
of homogeneity in
$\eta^\mu$. The component fields themselves $F^{_{(+)}}_{n} (p,
\eta)$ and $F^{_{(-)}}_{n} (p, \eta)$ describe states with positive
and negative helicities $\lambda=n$ and $\lambda=-n$.

The explicit form of the functions (\ref{rwd8a}) reproduces
automatically the equations of motion of the field $F_{n} (p,
\eta)$:
\begin{eqnarray}
\lb{rwd9}
p^2 \, F_{n} (p, \eta)   &=&  0 \, , \\ [7pt]
\lb{rwd10}
\left(p \cdot \frac{\partial}{\partial \eta}\right) \, F_{n} (p, \eta) &=& 0 \, , \\ [7pt]
\lb{rwd11}
\left(\frac{\partial}{\partial \eta} \cdot \frac{\partial}{\partial \eta}\right) \, F_{n} (p, \eta) &=& 0 \, , \\ [7pt]
\lb{rwd12}
\left(\eta \cdot \frac{\partial}{\partial \eta}\right) \, F_{n} (p, \eta) &=& n \, F_{n} (p, \eta)  \, .
\end{eqnarray}
The last equation determines the degree of homogeneity for the field
$F_{n} (p, \eta)$ in the variables $\eta^\mu$. In addition, the
presence in the definition of $\Psi_{n} (p, \eta)$ of the field
$F_{n} (p, \eta)$ together with the $\delta$-function $\delta (\eta
\cdot p)$ leads to the following equivalence relation:
\be
\lb{rwd13}
F_{n} (p, \eta) \ \ \sim \ \ F_{n} (p, \eta)  + (p \cdot
\eta) \, \epsilon_{n-1} (p,\eta) \, ,
\ee
where the functions
$\epsilon_{n-1} (p,\eta)$ satisfy equations (\ref{rwd9}) --
(\ref{rwd11}) and have a degree of homogeneity $(n-1)$ with respect
to the variable $\eta$. Relation (\ref{rwd13}) is essentially a
gauge transformation with parameters $\epsilon_{n-1} (p,\eta)$ and,
therefore, the field $F_{n} (p, \eta)$ is a gauge field\footnote{The
description of massless gauge fields using distributions was
considered, for example, in \cite{BeM,BekSk}.}.

The standard tensor description of gauge fields is obtained after
explicit selecting the polynomial dependence in $\eta$ of the
field $F_{n} (p, \eta)$:
\be
\lb{eta-exp}
F_{n} (p, \eta)= \eta^{\mu_1}\ldots\eta^{\mu_n} \, f_{\mu_1\ldots\mu_n}(p)  \,
\ee
and transferring to the coordinate representation. The corresponding
coordinate tensor field $f_{\mu_1\ldots\mu_n}(x)$ is automatically
totally symmetric with respect to the vector indices
$f_{\mu_1\ldots\mu_n}(x)=f_{(\mu_1\ldots\mu_n)} (x)$ and, thanks to
(\ref{rwd9})-(\ref{rwd11}), obeys the equations
\be
\lb{f-eq}
\Box
f_{\mu_1\ldots\mu_n}(x)=0\,,\qquad
\partial^{\mu_{1}}f_{\mu_1\ldots\mu_n}(x)=0\,, \qquad
\eta^{\mu_1\mu_2}f_{\mu_1\mu_2\ldots\mu_n}(x)=0 \, .
\ee
In addition, the equivalence relation (\ref{rwd13}) means that the
fields $f_{\mu_1\ldots\mu_n}(x)$ are defined up to the gauge
transformations:
\be
\lb{f-ga-tr} \delta
f_{\mu_1\mu_2\ldots\mu_n}(x)=\partial_{(\mu_1}\epsilon_{\mu_2\ldots\mu_n)}(x)
\, .
\ee
 Equations (\ref{f-eq}) and gauge symmetry (\ref{f-ga-tr})
are standard conditions that define free massless higher spin fields.

In the case of zero helicity, the use of additional variables
$\eta^\mu$ is not required. In this case, the relativistic field
coincides with the Wigner wave function $\Psi_0(p)=\Phi_0(p)$, which is
not a gauge field, and obeys only the Klein-Gordon equation in the
momentum representation: $p^2\Psi_0(p)=0 $.

\setcounter{equation}{0}
\section{Conclusion}

We have proposed a generalization of the Wigner operator and, on its
basis, constructed a $4D$ relativistic field description of the
irreducible massless representations of the extended Poincar\'e
group that complements the proper Poincar\'e group with discrete
transformations $P$, $C$, $T$. In contrast to the standard
irreducible local field representations, which are formulated  in
terms of field strengths with fixed helicities (see, for example,
\cite{Weinb2}, \cite{BK}), the field irreducible representations of
the extended Poincar\'e group that we have constructed are
formulated in terms of gauge potentials satisfying the conditions
defining free massless higher spin fields.

We have constructed a local field description in terms of the fields
$\Psi(p,\eta)$ in the momentum representation depending on an
additional vector variable $\eta^{\mu}$ and describing two massless
states with helicities $\lambda$ and $ -\lambda$. Equations
(\ref{th-0})\,,(\ref{b-0}) for the function $\Psi(p,\eta)$ were
obtained and their solution (\ref{rwd7}) was found. A remarkable
property of this solution is the presence of the generalized
function $\delta(\eta \cdot p)$ as a multiplier:
$\Psi(p,\eta)=\delta(\eta \cdot p)F(p,\eta)$ . This automatically
leads to the gauge invariance of the field $F(p,\eta)$.

We note that earlier a similar approach for describing infinite spin
irreducible representations of the Poincar\'e group based on the
generalized Wigner operator was developed in our paper \cite{BIPF} (see also \cite{ShTor1}, \cite{ShTor2}).
Here we have shown that this approach can also be successfully
applied to the description of irreducible massless representations
of the extended Poincar\'e group with finite integer helicities and,
thus, we have demonstrated a certain universality of this approach.
It seems interesting to construct local relativistic fields with
finite half-integer helicities based on the generalized Wigner
operator, which we plan to do in the forthcoming papers.

\smallskip
\section*{Acknowledgment}
The work of I.L.B. is supported by the Ministry of Education of the
Russian Federation, project No. QZOY-2023-0003. The work of S.A.F.
is supported by RSF grant No. 21-12-00129.

\section*{Appendix\,1: \\
Covering  of the $4D$ Lorentz group. Wigner operators.}
\def\theequation{A.\arabic{equation}}
\setcounter{equation}0

The relationship between the Lorentz group and its covering group
$SL(2,\mathbb{C})$ is formulated in the frame work of space
$\mathcal{H}$ of Hermitian $(2\times2)$ matrices. In this case the
basis matrices are $\sigma^0 =I_2$ and $\sigma^i$, $i=1,2,3$ are the
$\sigma$-Pauli matrices. These matrices establish a one-to-one
correspondence between the set of vectors in the space
$x_{\mu}\in\mathbb{R}^{1,3}$ and the set $\mathcal{H}$ of Hermitian
matrices $X=x_{\mu} \sigma^{\mu} \in \mathcal{H}$. The action of the
group $SL(2,\mathbb{C})$ on the set $\mathcal{H}$
\be
\lb{sl-h-p} X
\to X' = A X A^{\dagger}\, , \;\;\; X, X' \in \mathcal{H}\, , \;\;\;
A \in SL(2,\mathbb{C})
\ee
leads to the group homomorphism
$SL(2,\mathbb{C}) \to SO^{\uparrow}(1,3)$ represented in
(\ref{Alambd-p}).

As a test 4-momentum of a massless particle, we take the vector $\overset{_{\mathrm{\;o}}}{p} \, \in \,{\mathbb{R}^{1,3}}$, which has the following components
\be \lb{fr3-p}
||\overset{_{\mathrm{\;o}}}{p}_{\mu}|| = (\overset{_{\mathrm{\;o}}}{p}_{0}, \overset{_{\mathrm{\;o}}}{p}_{1},
\overset{_{\mathrm{\;o}}}{p}_{2},
\overset{_{\mathrm{\;o}}}{p}_{3})= (E,0,0,E) \, .
\ee
The Wigner operator $A_{(p)} \in SL(2, \mathbb{C})$ is defined by the matrix equation
\be \lb{fr4-p}
A_{(p)} (\overset{_{\mathrm{\;o}}}{p}\,  \sigma) A_{(p)}^{\dagger} = (p\,\sigma) \, ,
\ee
where $(x \sigma) : = x_{\mu} \sigma^{\mu}$.
The arbitrariness in the definition of Wigner operators is fixed by the equality $A_{(\overset{_{\mathrm{\;o}}}{p} )} = I_2$. Relation (\ref{fr4-p}) in vector representation has the form
$ \Lambda_\nu^{\;\; \mu}(A_{(p)}) \, \overset{_{\mathrm{\;o}}}{p}_{\mu} = p_{\nu}$. The matrices $h \in SL(2,\mathbb{C})$ from the stability subgroup $G_{\overset{_{\mathrm{\;o}}}{p}}$ preserve the test momentum,
\be \lb{fr5-p}
h \; (\overset{_{\mathrm{\;o}}}{p}\, \sigma) \; h^{\dagger} = (\overset{_{\mathrm{\;o}}}{p}\, \sigma) \,.
\ee
In the case of an isotropic 4-momentum (\ref{fr3-p}), the elements $h \in G_{\overset{_{\mathrm{\;o}}}{p}}\simeq ISO(2)$ have the form
\be
\lb{hhhh}
h =  \begin{pmatrix} 1 & b_1 + {\sf i} b_2 \\ 0 & 1 \end{pmatrix} \,
\begin{pmatrix} e^{\frac{i}{2} \theta} & 0  \\ 0 & e^{-\frac{i}{2} \theta}
\end{pmatrix}\, ,
\ee
i.e., the matrix $h$ depends on three parameters $\theta \in [0,2\pi]$ and $\vec{b} =(b_1,b_2)
\in \mathbb{R}^2$.

The Wigner operator $A_{(p)}$ is defined up to right
multiplication by an element from the stability subgroup
$G_{\overset{_{\mathrm{\;o}}}{p}}$ and thus parameterizes the coset
space $SL(2, \mathbb{C})/G_{\overset{_{\mathrm{\;o}}}{p}}$. The
relation $A \, A_{(p)} = A_{(\Lambda p)} \, h_{A,p}$ defining the
action of the element $A \in SL(2, \mathbb{C})$ on the  coset space
$SL(2, \mathbb{C})/G_{\overset{_{\mathrm{\;o}}}{p}}$ parameterized
by the Wigner operator leads to two consequences
\be
\lb{acgf1-p}
h_{A,p} = A_{(\Lambda p)}^{-1} \, A \, A_{(p)} \;\;\;\; \Rightarrow
\;\;\;\; h_{A,\Lambda^{-1} p} = A_{(p)}^{-1} \, A \,
A_{(\Lambda^{-1} p)} \, ,
\ee
where the indices at $h_{A,p}$
indicate that the element of the test momentum stability subgroup
depends on $A \in SL(2,\mathbb{C})$ and the 4-momentum $p$.

The parameters that correspond to the stability subgroup element
$h_{A,p}$ specified in (\ref{acgf1-p}) are denoted by $\theta_{A,p}$
and $\vec{b}_{A, p}$ and are determined from the relation
\be
\lb{acgf1-p1}
h_{A,p} = A_{(\Lambda
p)}^{-1} \, A \, A_{(p)} =  \begin{pmatrix} 1 & \mathbf{b}_{A,p} \\
0 & 1 \end{pmatrix} \, \begin{pmatrix} e^{\frac{i}{2} \theta_{A,p}}
& 0  \\ 0 & e^{-\frac{i}{2} \theta_{A,p}} \end{pmatrix}\, ,
\ee
where $\mathbf{b}_{A,p}=(b_{A,p})_1 + i (b_{A,p})_2$. The parameters
$\theta_{A,\Lambda^{-1} p}$ and $\vec{b}_{A, \Lambda^{-1} p}$
corresponding to the second relation from (\ref{acgf1-p}), are
determined from $\theta_{A,p}$ and $\vec{b}_{A,p}$ by the
substitution: $p \to \Lambda(A)^{-1} p$. Taking into account
(\ref{acgf1-p}), the condition $A_{(\overset{_{\mathrm{\;o}}}{p})} =
I_2$ leads to the equalities: \be \lb{add-re-p}
 \theta_{A_{(p)}, \overset{_{\mathrm{\;o}}}{p}} = 0 \, , \;\;\; \vec{b}_{A_{(p)}, \overset{_{\mathrm{\;o}}}{p}} = 0 \, .
\ee

\section*{Appendix\,2: \\
Pauli-Lubanski pseudovector for fields with an additional vector variable}
\def\theequation{B.\arabic{equation}}
\setcounter{equation}0

The Poincar\'e group generators $\hat{P}_\nu$, $\hat{M}_{\nu \mu}$, acting
in the field space $\Psi(p,\eta)$, have the form
\be \lb{genP}
\hat{P}_\mu = p_\mu \, , \;\;\; \hat{M}_{\mu \nu} = i\Bigl ( p_{\mu}
\frac{\partial}{\partial p^{\,\nu}} - p_{\nu} \frac{\partial}{\partial
p^{\,\mu}} + \eta_{\mu} \frac{\partial}{\partial \eta^{\,\nu}} - \eta_{\nu}
\frac{\partial}{\partial \eta^{\,\mu}} \Bigr ) \, .
\ee
The
Pauli-Lubanski pseudovector in the reference frame of a test
momentum is defined as follows:
\be
\lb{pvPL1}
\overset{_{\mathrm{\;o}}}{W}_\mu = \frac{1}{2}
\varepsilon_{\mu\nu\lambda\rho} \hat{M}^{\lambda \rho}
{\overset{_{\mathrm{\;o}}}{p}}{}^\nu = i
\varepsilon_{\mu\nu\lambda\rho} \,
{\overset{_{\mathrm{\;o}}}{p}}{}^\nu \,\eta^{\lambda}
\frac{\partial}{\partial \eta_{\rho}} \, .
\ee
Its components have
the following form: \be \lb{pvPL2}
\begin{array}{rcl}
\displaystyle
\overset{_{\mathrm{\;o}}}{W}_0 = \overset{_{\mathrm{\;o}}}{W}_3 &=& \displaystyle
E \left( \zeta \frac{\partial}{\partial \zeta} - \bar{\zeta} \frac{\partial}{\partial \bar{\zeta}} \right) \, , \\[0.5cm]
\displaystyle
\overset{_{\mathrm{\;o}}}{W}_{(+)} =- \frac{1}{2} \left (\overset{_{\mathrm{\;o}}}{W}_1+ i  \overset{_{\mathrm{\;o}}}{W}_2 \right )  &=&
\displaystyle E\left( \eta^+\frac{\partial}{\partial \zeta} + \bar{\zeta} \frac{\partial}{\partial \eta^{-}} \right ) \, ,  \\[0.5cm]
\displaystyle
\overset{_{\mathrm{\;o}}}{W}_{(-)} =- \frac{1}{2} \left (\overset{_{\mathrm{\;o}}}{W}_1- i  \overset{_{\mathrm{\;o}}}{W}_2 \right ) &=&
\displaystyle
E \left ( \eta^+\frac{\partial}{\partial \bar{\zeta}} + \zeta \frac{\partial}{\partial \eta^{-}} \right )\, ,
\end{array}
\ee
where the variables defined in (\ref{nev1-p}) are used.
It can be shown that in an arbitrary reference frame the components of the Pauli-Lubanski vector are given by the expressions:
\be \lb{pvPL3}
\begin{array}{c}
\displaystyle
W_0 = \, - \frac{1}{2} \, \left[ (\varepsilon_{(+)} \cdot \eta) \left(\varepsilon_{(-)} \cdot \frac{\partial}{\partial \eta}\right)- (\varepsilon_{(-)} \cdot \eta) \left (\varepsilon_{(+)} \cdot \frac{\partial}{\partial \eta}\right) \right ] = W_3  \, ,  \\[0.5cm]
\displaystyle
W_{(\pm)} = \left ( (\varepsilon_{(\mp)} \cdot \eta) (p \cdot \frac{\partial}{\partial \eta}) - (p \cdot \eta) (\varepsilon_{(\mp)} \cdot \frac{\partial}{\partial \eta}) \right).
\end{array}
\ee

\section*{Appendix\,3: \\
Polarization vectors}
\def\theequation{C.\arabic{equation}}
\setcounter{equation}0

Here we give expressions for the 4-polarization vectors used in the
paper.

In the standard momentum frame, the 4-polarization vectors that are
orthogonal to each other and to the massless test momentum
$\overset{_{\mathrm{\;o}}}{p}$,
 have components
 \be
 \lb{sol2-p}
(\overset{_{\mathrm{\,o}}}{\varepsilon}_{(1)})_\nu = (0,1,0,0) \, ,
\qquad (\overset{_{\mathrm{\,o}}}{\varepsilon}_{(2)})_\nu =
(0,0,1,0) \, .
\ee
In an arbitrary basis, they are denoted by
\be
\lb{pv-11-p}
\begin{array}{c}
\varepsilon_{(1)} := \Lambda( A_{(p)}) \, \overset{_{\mathrm{\,o}}}{\varepsilon}_{(1)} \, ,\;\;\; \varepsilon_{(2)} := \Lambda( A_{(p)}) \, \overset{_{\mathrm{\,o}}}{\varepsilon}_{(2)} \, .
\end{array}
\ee
The following linear combinations of vectors (\ref{pv-11-p}) are also used:
\be \lb{pv-p}
\varepsilon_{(\pm)} :=  \varepsilon_{(2)} \pm i  \varepsilon_{(1)} \, ,
\ee
which, by construction, satisfy the equations
\be \lb{pr-vp-p}
p \cdot \varepsilon_{(\pm)} = 0, \qquad \varepsilon_{(\pm)}  \cdot \varepsilon_{(\pm)}  = 0 \, , \qquad \varepsilon_{(+)} \cdot \varepsilon_{(-)} = -2 \, .
\ee


\begin{thebibliography}{90}


\bibitem{Wig39}
E.P.\,Wigner,
{\it On Unitary Representations of the Inhomogeneous Lorentz Group},
Annals Math. {\bf 40} (1939), 149.

\bibitem{Wig48}
E.P.\,Wigner,
{\it Relativistische Wellengleichungen},
Z. Physik {\bf 124} (1948) 665.

\bibitem{BarWig48}
V.\,Bargmann, E.P.\,Wigner,
{\it Group Theoretical Discussion of Relativistic Wave Equations},
Proc. Nat. Acad. Sci. {\bf 34} (1948), 211.

\bibitem{FG} L.\,Fonda, G.C.\,Ghirardi, {\it Symmetry Principles in Quantum Physics}, N.Y., Marcel Deccer, INC, 1970, 515 p.

\bibitem{Nov} Yu.V.\,Novozhilov, {\it Introduction to Elementary Particle Theory}, Volume 78 in International Series in Natural Philosophy, Pergamon Press, Oxford, 1975.

\bibitem{BR}
A.O.\,Barut, R.\,Raczka, {\it Theory of group representations and
applications,} World Scientific Publishing Company, 1986.

\bibitem{TUNG} Wu-Ki\,Tung, {\it Group Theory in Physics}, World Scientific, 1985, 344 p.

\bibitem{BekSk}
X.\,Bekaert, E.D.\,Skvortsov, {\it Elementary particles with
continuous spin}, Int. J. Mod. Phys.   {\bf A32} (2017) 1730019,
{\tt arXiv:1708.01030\,[hep-th]}.

\bibitem{Vilen}
N.Y.\,Vilenkin, {\it Special functions and the theory of group representations,} American Mathematical Soc., 1978.

\bibitem{ZhSh}
D.P.\,Zhelobenko, A.I.\,Shtern, {\it Representations of Lie groups,} Nauka, Moscow, 1983 (in Russian).

\bibitem{BIPF}
I.L\,Buchbinder, A.P.\,Isaev, M.A.\,Podoinitsyn, S.A.\,Fedoruk, {\it Generalization of the Bargmann-Wigner construction for infinite spin fields}, {\tt arXiv:2303.11852\,[hep-th]}.

\bibitem{ZFed}
S.A.\,Fedoruk, V.G.\,Zima, {\it Covariant quantization of d = 4 Brink-Schwarz superparticle with Lorentz harmonics,} Theor. Math. Phys. {\bf 102} (1995) 305–322

\bibitem{Weinb1} S.\,Weinberg,  {\it Feynman rules for any spin}, Phys. Rev. {\bf 133}(5B) (1964) B1318.

\bibitem{Weinb2} S.\,Weinberg, {\it Feynman rules for any spin. II. Massless particles}, Phys. Rev. {\bf 134}(4B) (1964) B882.

\bibitem{BK}
I.L.\,Buchbinder, S.M.\,Kuzenko, {\it Ideas and Methods of Supersymmetry and Supergravity Or a Walk Through Superspace,} IOP Publishing, Bristol and Philadelphia, 1998.

\bibitem{GS}
D.M.\,Gitman, A.L.\,Shelepin, {\it Fields on the Poincare group: arbitrary spin description and relativistic wave equations}, Int.J.Theor.Phys. {\bf 40} (2001) 603-684, {\tt arXiv:hep-th/0003146}.

\bibitem{BGS}
I.L.\,Buchbinder, D.M.\,Gitman, A.L.\,Shelepin, {\it Discrete symmetries as automorphisms of the proper Poincare group},  Int.J.Theor.Phys. {\bf 41} (2002) 753-790, {\tt arXiv:hep-th/0010035}.

\bibitem{BeM} X.\,Bekaert, J.\,Mourad, {\it The continuous spin limit of higher spin field equations}, JHEP {\bf 0601}
(2006) 115, {\tt arXiv:hep-th/0509092}.

\bibitem{ShTor1}
P.\,Schuster, N.\,Toro,
{\it On the theory of continuous-spin particles: wavefunctions and soft-factor scattering amplitudes},
JHEP {\bf 09} (2013) 104, {\tt arXiv:1302.1198\,[hep-th]}.

\bibitem{ShTor2}
P.\,Schuster, N.\,Toro,
{\it On the theory of continuous-spin particles: helicity correspondence in radiation and forces},
JHEP {\bf 09} (2013) 105, {\tt arXiv:1302.1577\,[hep-th]}.

\bibitem{IsPod0} A.P.\,Isaev, M.A.\,Podoinitsyn,  {\it Two-spinor description of massive particles and relativistic spin projection operators,} Nucl. Phys. B {\bf 929} (2018) 452, {arXiv:1712.00833 [hep-th]}.







































\end{thebibliography}
\end{document}